%% file: RJwrapper.tex
\documentclass[a4paper]{report}
\usepackage[utf8]{inputenc}
\usepackage[T1]{fontenc}
\usepackage{RJournal}
\usepackage{amsmath,amssymb,array}
\usepackage{booktabs}
\usepackage{url}
\usepackage{float}

\begin{document}
	
	%% do not edit, for illustration only
	\sectionhead{Contributed research article}
	\volume{XX}
	\volnumber{YY}
	\year{20ZZ}
	\month{AAAA}
	
	%% replace RJtemplate with your article
	\begin{article}
		\input{filename}

	\end{article}
	
\end{document}

%% file: filename.tex
\title{A framework for estimating and visualising excess mortality during the COVID-19 pandemic}

\author{by G. Konstantinoudis, V. G\'{o}mez-Rubio, M. Cameletti,  M. Pirani, G. Baio, M. Blangiardo}

\maketitle

\abstract{
COVID-19 related deaths underestimate the pandemic burden on mortality because they suffer from completeness and accuracy issues. Excess mortality is a popular alternative, as it compares observed with expected deaths based on the assumption that the pandemic did not occur. Expected deaths had the pandemic not occurred depend on population trends, temperature, and spatio-temporal patterns. In addition to this, high geographical resolution is required to examine within country trends and the effectiveness of the different public health policies. In this tutorial, we propose a framework using R to estimate and visualise excess mortality at high geographical resolution. We show a case study estimating excess deaths during 2020 in Italy. The proposed framework is fast to implement and allows combining different models and presenting the results in any age, sex, spatial and temporal aggregation desired. This makes it particularly powerful and appealing for online monitoring of the pandemic burden and timely policy making.
}

\section{Introduction}

Estimating the burden of the COVID-19 pandemic on mortality is an important challenge~\citep{weinberger2020estimation}. COVID-19 related deaths are subject to testing capacity and changes in definition and reporting, raising accuracy and completeness considerations \citep{aburto2021estimating, konstantinoudis2021regional}. In addition, COVID-19 related deaths give an incomplete picture of the burden of the COVID-19 pandemic on mortality, as they do no account for the indirect pandemic effects due to, for instance, disruption to health services~\citep{kaczorowski2021beyond}. Alternatively, excess mortality has been extensively used to evaluate the impact of the COVID-19 pandemic on mortality \citep{rossen2020excess, islam2021excess, kontis2020magnitude, konstantinoudis2021regional, VerbeeckLMMCOVID}.

Excess mortality is estimated by comparing the observed number of deaths in a particular time period with the expected number of deaths under the counterfactual scenario that the pandemic did not occur. Typically this counterfactual scenario is calculated using a comparison period, for instance several previous years (\url{https://www.euromomo.eu/}). Calculating accurately the expected deaths requires accounting for factors such as population trends, seasonality, temperature, public holidays and spatio-temporal dependencies. While accounting for the above-mentioned factors, most studies to date have estimated excess mortality at national level \citep[e.g.,][]{rossen2020excess, weinberger2020estimation} and a few have looked across different countries reporting important geographical discrepancies~\citep{islam2021excess, kontis2020magnitude, kontis2021lessons}. 

While national level estimates of excess mortality give valuable insights into the total number of excess deaths, they do not allow the evaluation of within country geographical disparities. Such information is essential to understand the country's transmission patterns and effectiveness of local policies and measures to contain the pandemic \citep{kontopantelis2021excess}. Temporal trends in excess can substantially differ across regions of the same country, which makes national-based comparisons even more challenging \citep{Blangiardo2020}. Therefore, understanding the burden of the COVID-19 pandemic on mortality requires higher spatial resolution and models that account for spatial, temporal and spatio-temporal dependencies. 

When working at high spatio-temporal resolution, data are generally sparse, leading to excess mortality estimates that are highly variable. This is aggravated by the fact that they are subject to spatial and temporal correlation, making it essential to use statistical methods that account for these dependencies in order to provide robust and accurate estimates. The disease mapping framework, which is commonly employed in epidemiology to study the spatio-temporal variation of diseases \citep{waller1997hierarchical, moragarjournal}, can be exploited to estimate excess mortality at subnational and weekly level. The Bayesian approach is naturally suited in this context, as it is able to model complex dependency structures, as well as to incorporate uncertainty in the data and modelling process. In addition, while fully propagating the uncertainty, it allows us to summarise the results at any desired spatio-temporal aggregation (using for instance coarser geographical units suitable for policy implementation). This in combination with fast approximate methods to inference, such as the Integrated Laplace Approximation (INLA, \citep{rue2009approximate}), make this framework particularly powerful and appealing for monitoring of the pandemic burden and timely policy making.

\begin{figure}[ht]
	\centering
	\includegraphics{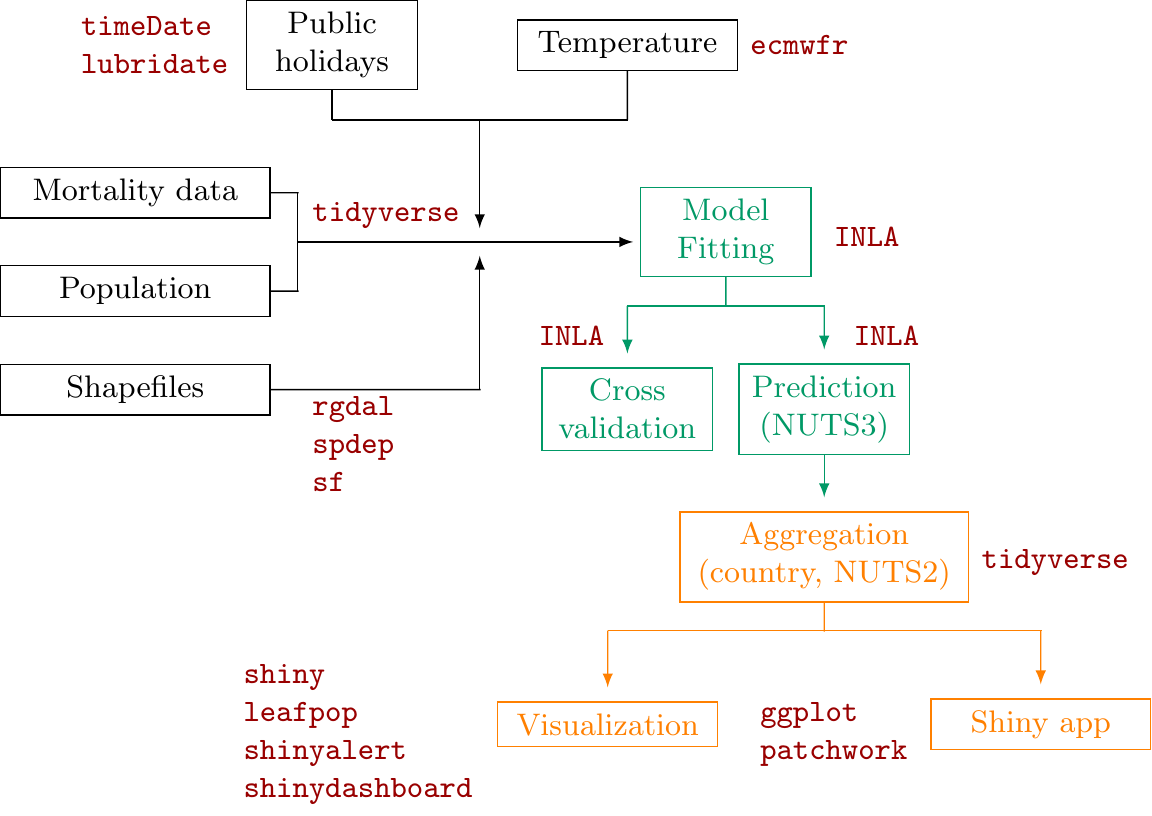}
	\caption{Diagram of the workflow: data wrangling in black, analysis in green, post-processing in orange and main packages in red. NUTS stands for Nomenclature of Territorial Units for Statistics with NUTS3 being the highest spatial resolution available and NUTS2 coarser but appropriate for policy making.}
	\label{fig:workflow}
\end{figure}

In this tutorial, we describe how to run a study on excess mortality at high spatial and temporal resolution using a Bayesian approach and R. This tutorial provides a detailed implementation of the approach followed previously in 5 European regions \citep{konstantinoudis2021regional} with a slightly modified main model \citep{riou2023direct}. Figure \ref{fig:workflow} illustrates graphically the workflow followed in this paper together with the main r-packages used. Source code for replicating the data wrangling (R-files starting with 01, 02 or 03), analysis (R-files starting with 04) and post-processing steps (R-files starting with 05 or 06 and the App folder) and data files are available from GitHub at \url{https://github.com/gkonstantinoudis/TutorialExcess}.

This tutorial is structured as follows: we first describe the modelling framework and present the case study in Italy. We then show how to run and evaluate the model, and extract and visualise the results. Finally, we present an R-shiny app which makes the results effectively and easily disseminated.

\section{Bayesian hierarchical spatio-temporal model to estimate excess mortality}

We propose a Bayesian hierarchical model to quantify spatio-temporal excess mortality under extreme events such as the COVID-19 pandemic, stratified by specific age-sex population groups. To do so, we first describe the statistical model for predicting the number of deaths from all-causes based on historical data, in the counterfactual scenario in which the pandemic did not take place. Then, we show how to estimate the magnitude of excess deaths over a specific period of time, with associated uncertainty, by comparing the predicted versus the actual number of deaths.

Let $y_{jstk}$ and $P_{jstk}$ be the number of all-cause deaths and the population at risk for the $j$-th week, $j=1, \dots, J_t$, where $J_t$ is the total number of weeks of the year $t$ ($t=2015, \dots 2019$), the $s$-th spatial unit ($s=1, \dots S$, where $S$ is the number of provinces in Italy), and $k$-th age-sex group ($k = 1, \dots 10$). Also let $x_{1jt}$, $x_{2t}$ and $z_{jst}$ denote the adjustment covariates, respectively public holidays (i.e., $x_j$ = 1 if week $j$ contains a public holiday and 0 otherwise), the year that the $j$-the week belongs to and temperature. We assume that the historical observed number of deaths, conditional on the risk $r_{jstk}$, follows a Poisson distribution, with a log-linear model on the risk. To simplify notation, we omit $k$, although the following model was fitted to all $k$ age-sex groups:

\begin{align} \label{eq:1}
\begin{split}
y_{jst}|r_{jst} \sim & \text{Poisson}\big(\mu_{jst}=r_{jst}P_{jst}\big),  \\
\log \left(r_{jst} \right) & = \beta_{0jt} + \beta_{1} x_{1jt} + \beta_{2} x_{2t} + f(z_{jst}) + b_{s} + w_{jt}.
\end{split}
\end{align}

Here, $\beta_{0jt}$ is the week specific intercept in year $t$ given by $\beta_{0jt}=\beta_{0}+\epsilon_{jt}$ for the $k$-th age-sex group, where $\beta_{0}$ is the global intercept and $\epsilon_{jt}$ is an unstructured random effect representing the deviation of each week from the global intercept, which is modelled using independent and identically (iid) distributed Gaussian prior distribution with zero-mean and variance equal to $\tau_\epsilon^{-1}$. The parameter $\beta_1$ and $\beta_2$ are  unknown regression coefficients associated to the public holidays covariate $x_{1jt}$ and a long term linear trend. The effect of temperature is allowed to be non-linear $f(\cdot)$ by specifying  a second-order random walk (RW2) model:

\begin{equation*}
z_{jst} \mid z_{(j-1)st}, z_{(j-2)st}, \tau_z \sim \text{Normal}\left(2z_{(j-1)st}+z_{(j-2)st},\tau_z^{-1}\right),
\end{equation*}

\noindent$\tau_z^{-1}$ is the variance.

Terms $b_s$ and $w_j$ are spatial and temporal random effects, respectively. We specify the spatial random effect term, $b_s$, with a convolution prior \citep{besag1991bayesian}, and the temporal random effect term, $w_j$ with a non-stationary in time prior. In detail, we model $\boldsymbol{b}$ using a reparameterisation of the popular Besag-York-Molli\'e prior, which is a convolution of an intrinsic conditional autoregressive (CAR) model and an iid Gaussian model. Let $u_s$ be the spatially structured component defined by an intrinsic CAR \citep{Besag1974} prior  $u_s|u_i,i\in \partial_s \sim(\bar{u},\tau_u^{-1}/m_s)$, where $\bar{u}$ is the local mean and $\partial_s$ and $m_s$ are respectively the set and the number of neighbours of area $s$,  $\tau_u^{-1}$ the conditional variance, and $v_s$  the unstructured component with prior $v_s \overset{iid}{\sim} \text{Normal}(0,\tau_v^{-1})$. We model $b_s$ as follows \citep{besag1991bayesian, riebler2016intuitive,  konstantinoudis_discrete}:
\[
b_s=\frac{1}{\sqrt{\tau_b}}\left(\sqrt{1-\phi}v^\star_s+\sqrt{\phi}u^\star_s\right)
\]
where $u_s^\star$ and $v_s^\star$ are standardised versions of $u_s$ and $v_s$ to have variance equal to 1 \citep{simpson2017penalising}.
The term $0\leq \phi\leq 1$ is a mixing parameter, which measures the proportion of the marginal variance explained by the structured effect.
Finally, we assign to the temporal random effect term, $w_{jt}$, a Gaussian random walk model of order 1 (RW1). This component captures seasonality and is specified as:
\[
w_{jt} \mid w_{(j-1)t}, \tau_w \sim \text{Normal}(w_{(j-1)t},\tau_w^{-1}).
\]

The Bayesian representation of the above model is completed once we select priors for the fixed effects $\beta_0$ and {\boldmath$\beta$} and the hyperparameters: $\tau_{\epsilon}, \tau_z, \tau_b, \tau_w,$ and $\phi$. For the fixed effects we selected minimally informative Normal distributions whereas, we specified "penalising complexity" (PC) priors \citep{simpson2017penalising} for the hyperparameters. PC priors are defined by penalising deviations from a ``base'' model (e.g.,~specified in terms of a specific value of the hyperparameters) and have the effect of regularising inference, while not implying too strong information. Technically, PC priors imply an exponential distribution on a function of the Kullback–Leibler divergence between the base model and an alternative model in which the relevant parameter is unrestricted. This translates to a suitable ``minimally informative'', regularising prior on the natural scale of the parameter.

In order to quantify the weekly excess mortality at sub-national level for specific age-sex population groups, we need to predict the number of deaths had COVID-19 not occurred. In Bayesian analysis, this can be performed by drawing random samples from the posterior predictive distribution (that is, the distribution of unobserved values conditional on the observed values from previous years). Specifically, letting $\pmb{\theta}$ be the model parameters, $\mathcal{D}$ be the observed data, and $y_{jst^{*}}$ be the count of deaths that we want to predict, we have:

\begin{equation}\label{eq:2}
p(y_{jst^{*}}\mid \mathcal{D}) = \int p(y_{jst^{*}}\mid\pmb{\theta}) p(\pmb{\theta}\mid\mathcal{D})d\pmb{\theta}.
\end{equation}

Operationally, we first generate random samples from the joint posterior marginal of the fitted linear predictor specified in equations (\ref{eq:1}) at the highest spatial resolution available (NUTS3 regions; Nomenclature of Territorial Units for Statistics 3 regions, \url{https://ec.europa.eu/eurostat/web/nuts/background/}). Successively, we use these to be in turn the mean parameter of a Poisson distribution, to obtain the predicted number of deaths, which represents the baseline number of deaths assuming the pandemic did not take place. 

Finally, to estimate the magnitude of excess deaths, the predicted number of deaths is compared against the actual observed number of deaths, allowing the computation of the relative change in the mortality (i.e., relative to what we could expect if the pandemic did not occur). This is obtained by (i) subtracting the predicted number of deaths from the observed number of deaths in each time point $j$ in the  $t^*$-th year and spatial unit $s$ (number of excess deaths or NED), and (ii) dividing NED by the predicted number of deaths for each sample and multiplying by 100 (\% relative excess mortality or REM).

Bayesian inference for the model is computed using Integrated Nested Laplace Approximation (INLA; \citep{rue2009approximate}, which performs approximate Bayesian inference on the class of latent Gaussian models \citep{rue2005gaussian}. Unlike simulation based Markov chain Monte Carlo method, INLA is a deterministic algorithm, which employs analytical approximations and efficient numerical integration schemes to provide accurate approximations of the posterior distributions in short computing times. The INLA software is provided through the R package \texttt{INLA}, which can be downloaded from \url{https://www.r-inla.org/}. 

\section{Motivating example: Italy}

\subsection{Outcome data}
We retrieved all-cause mortality data during 2015-2020 in Italy from the Italian National Institute of Statistics (\url{https://www.istat.it/}). Data were available weekly (ISO week), by age (5-year age groups), sex and NUTS3 regions. As the COVID-19 mortality rates increase with  age, we aggregated mortality counts based on the following age groups: $<$40, 40-59, 60-69, 70-79 and 80 years and older \citep{davies2021community}.

\subsection{Population data}
Population data in Italy during 2015-2020 were retrieved from the Italian National Institute of Statistics. The data represent the population in Italy on January 1st of every year and are available by age (5-year age groups), sex and NUTS3 regions. To retrieve weekly population, we performed linear interpolation by the selected age groups ($<$40, 40-59, 60-69, 70-79 and 80$+$), sex and NUTS3 regions using populations at January 1st of the current and the next year. Population counts on January 1st 2021, which takes COVID-19 deaths in 2020 into consideration, were available at the time of analysis. Our goal was, however, to predict mortality for 2020 had the pandemic not occurred. Thus we performed an additional linear interpolation by age,  sex and NUTS3 regions to predict the population at January 1st 2021, using the years 2015-2020 (Figure~\ref{fig:popplot}, panel A). Object \texttt{pop} is a \texttt{tibble} containing the NUTS3 region ID (\texttt{NUTS318CD}), age group (\texttt{ageg}), sex (\texttt{sex}), year (\texttt{year}) and population counts (\texttt{population}):

\begin{figure}[!b]
	\centering
	\includegraphics{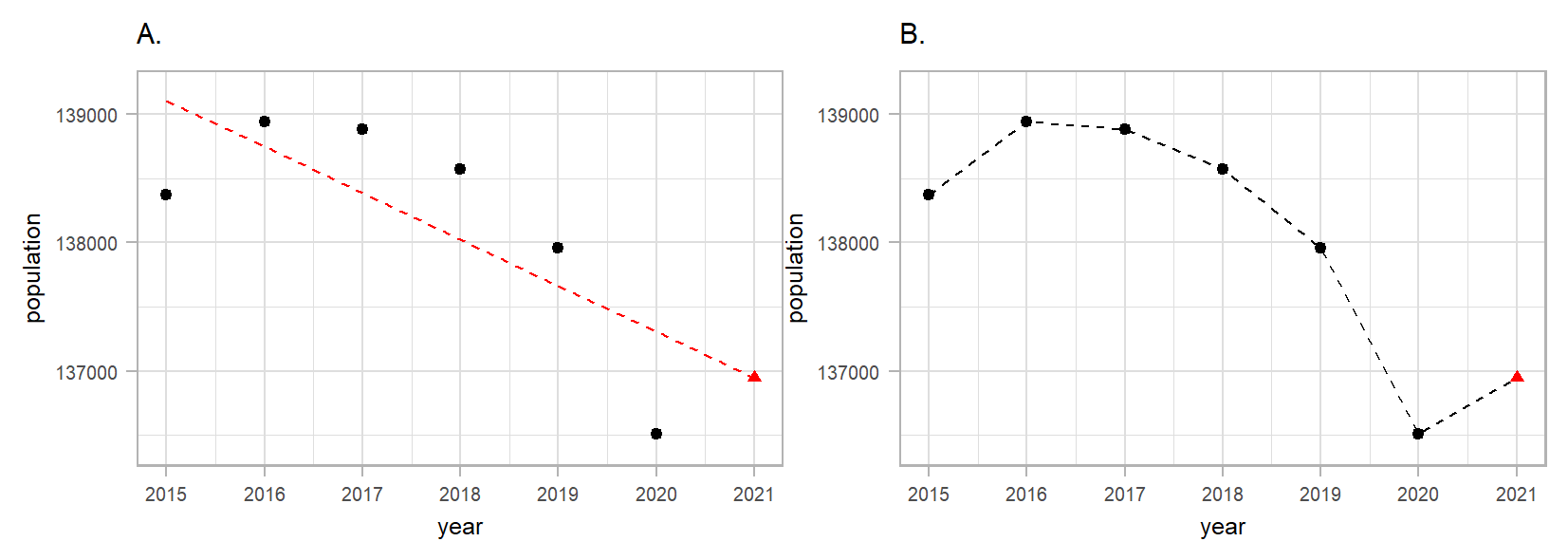}
	\caption{A schematic representation of the weekly population estimation procedure focusing on females aged 40-49 in Venice during 2015-2019 as an example. On panel A we show how we used the historical data (black points) and fit a linear regression (red dashed line) to predict 2021 (red triangle). On the panel B we show how we predicted weekly population by drawing lines between the years.}
	\label{fig:popplot}
\end{figure}

\begin{example}
pop
# A tibble: 6,420 x 5
# Groups:   ageg, sex [10]
NUTS318CD ageg   sex     year population
<chr>     <fct>  <chr>  <dbl>      <dbl>
1 TO        less40 female  2015     435758
2 TO        less40 female  2016     427702
3 TO        less40 female  2017     420498
4 TO        less40 female  2018     413141
5 TO        less40 female  2019     406937
6 TO        less40 female  2020     402768
7 TO        less40 male    2015     449605
8 TO        less40 male    2016     443941
9 TO        less40 male    2017     439522
10 TO        less40 male    2018     433365
# ... with 6,410 more rows
\end{example}

We can use the following code (based on \texttt{tidyverse} and "piping" principles) to calculate the population of the 1st of January 2021 by NUTS3 regions, sex and age:
\begin{example}
pop %>% group_by(NUTS3, sex, age) %>% 
	summarise(pop = as.vector(coef(lm(pop ~ year)) %*% c(1, 2021))) %>% 
	mutate(year = 2021) -> pop2021
\end{example}

We acknowledge that the linear trend in the population is a rather simplistic assumption. In subsequent analyses in Switzerland, we proposed a spatio-temporal approach similar with (\ref{eq:1}) to model the population counts had the pandemic not occurred \citep{riou2023direct}. The code for that analysis is also online available online (\url{https://github.com/jriou/covid19_ascertain_deaths}).\\
Once we obtained the year 2021 we performed an additional linear interpolation to calculate weekly number of population as shown on Figure~\ref{fig:popplot}, panel B.

\subsection{Covariates data}
We used covariates related with ambient temperature and national holidays and year of death to help the model predictions. Data on air-temperature during 2015-2020 in Italy at 2m above the surface of land were retrieved from the ERA5 reanalysis data set of the Copernicus climate change program \citep{hersbach2020era5}. The geographical resolution of the ERA5 estimates is 0.25$^\circ\times 0.25^\circ$ (panel A of Figure \ref{ERA5POINTS}). We calculated the weekly mean by the centroids of the 0.25$^\circ\times 0.25^\circ$ grid (panel B of Figure \ref{ERA5POINTS}) and then averaged the weekly temperature over the ERA5 centroids that overlay with the NUTS3 regions (panels B and C of Figure \ref{ERA5POINTS}). 

\begin{widefigure}[!b]
	\centering
	\includegraphics{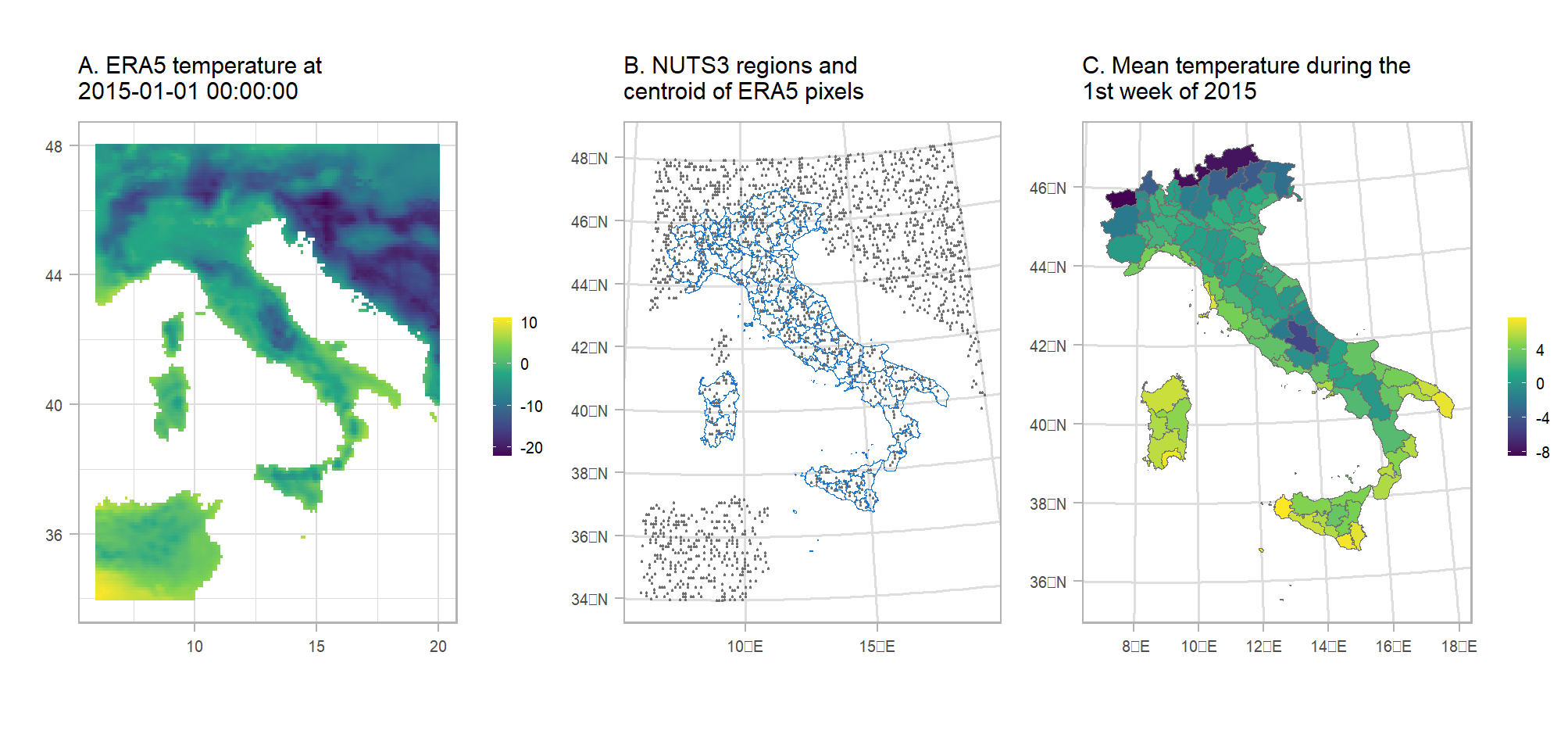}
	\caption{Schematic representation of the temperature misalignment procedure. A. The temperature obtained by ERA5 at 2015-01-01 00:00:00. B. NUTS3 regions in blue and a sample of the centroids of the pixels from the ERA5 raster. C. Mean temperature per NUTS3 region during the 1st week of 2015.}
	\label{ERA5POINTS}
\end{widefigure}

\subsection{Fitting the model}
The modelling process in \texttt{INLA} consists of three main steps: (1)~the selection of priors; (2)~definition of the model "formula" (which sets out the expression for the generalised linear predictor); and (3)~the call to the main function \texttt{inla}, which computes the estimates. 

\begin{figure}[t!]
	\centering
	\includegraphics{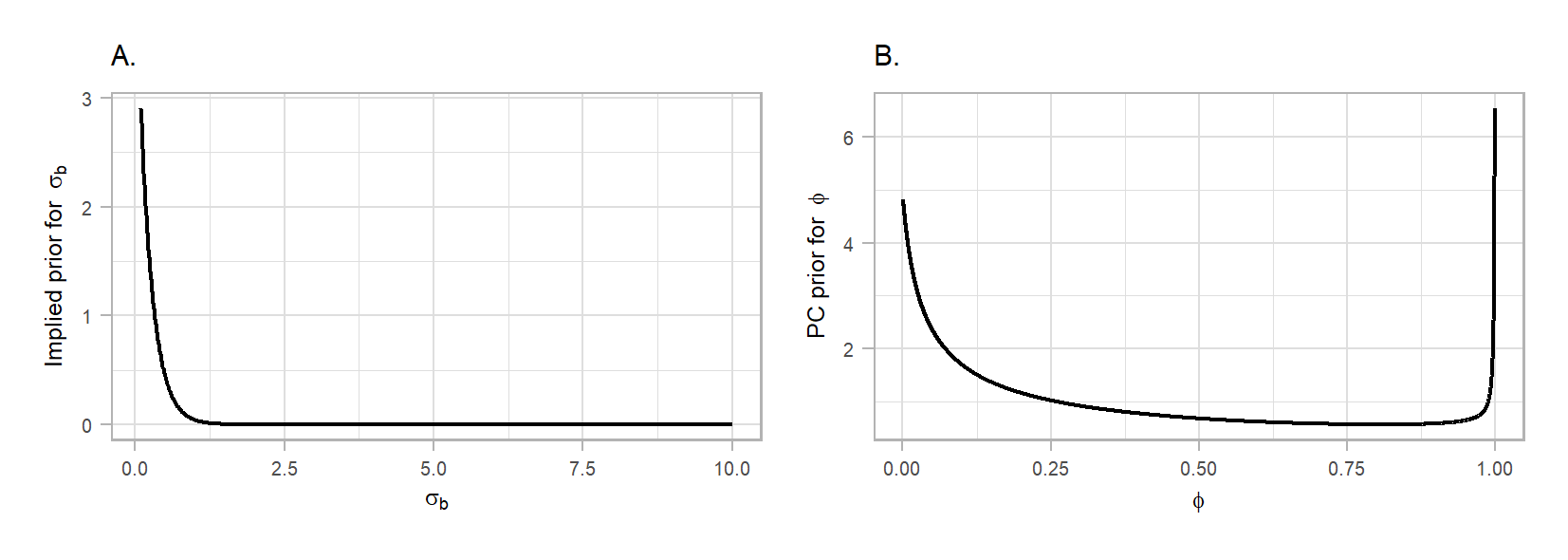}
	\caption{Penalised complexity (PC) priors for the hyperparameters of the spatial field. A. The implied PC prior for the standard deviation (as the original scale of the prior is the precision). B. The PC prior for the mixing parameter $\phi$.}
	\label{fig:priors}
\end{figure}

In particular, we constructed the PC priors for $\sigma_\epsilon=\sqrt{1/\tau_\epsilon}, \sigma_z=\sqrt{1/\tau_z}, \sigma_b=\sqrt{1/\tau_b}$ and $\sigma_w=\sqrt{1/\tau_w}$ based on statement that it is unlikely to have a relative risks higher than $\exp(2)$ based solely on spatial, yearly and seasonal variation, Figure~\ref{fig:priors}, panel A. For the mixing parameter $\phi$, we set $\text{Pr}(\phi<0.5) = 0.5$ reflecting our lack of knowledge about whether overdispersion or strong spatial autocorrelation should dominate the field~$b$, Figure~\ref{fig:priors}, panel B.

These assumptions can be encoded using the following code:
\begin{example}
# Defines the priors
hyper.bym <- list(
	theta1 = list('PCprior', c(1, 0.01)), 
	theta2 = list('PCprior', c(0.5, 0.5))
)
hyper.iid <- list(theta = list(prior="pc.prec", param=c(1, 0.01)))

# Defines the model "formula"
formula = 
	deaths ~ 1 + offset(log(population)) + hol + id.year + 
	f(id.tmp, model='rw2', hyper=hyper.iid, constr = TRUE, scale.model = TRUE) +
	f(id.wkes, model='iid', hyper=hyper.iid, constr = TRUE) + 
	f(id.time, model='rw1', hyper=hyper.iid, constr = TRUE, scale.model = TRUE,
	 cyclic = TRUE) +
	f(id.space, model='bym2', graph="W.adj", scale.model = TRUE, constr = TRUE, 
	hyper = hyper.bym)
	
control.family=inla.set.control.family.default()

# Calls INLA to fit the model
inla.mod = inla(formula,
	data=dat,
	family="Poisson",  
	verbose = TRUE, 
	control.family=control.family,
	control.compute=list(config = TRUE), 
	control.mode=list(restart=T),
	num.threads = round(parallel::detectCores()*.8), 
	control.predictor = list(link = 1))
\end{example}

After fitting the model, we take 1000 samples from the (approximated) posterior distribution of the linear predictor and we use each drawn sample as the mean of a Poisson distribution to retrieve the predicted mortality counts:

\begin{example}
post.samples <- inla.posterior.sample(n = 1000, result = inla.mod)
predlist <- do.call(cbind, lapply(post.samples, function(X)
exp(X$latent[startsWith(rownames(X$latent), "Pred")])))
	
pois.samples <- apply(predlist, 2, function(Z) rpois(n = length(Z), lambda = Z))
\end{example}

This allows us to estimate the entire predictive posterior distribution of the mortality counts incorporating both the sampling and the linear predictor uncertainty.

\subsection{Model validation}
To examine model validity, we performed a cross validation leaving out one historical year at a time and predicting the weekly number of deaths by NUTS3 regions for the year left out. For each stratum, we calculated the correlation between observed and fitted and a coverage probability, i.e. the probability that the observed death fall into the 95\% credible interval (95\% CrI) of the predicted. 

\section{Results}
\subsection{Cross validation}
We overall found good predictive ability of our model. The correlation estimates varies from 0.39 (95\% CrI: 0.37, 0.40) in females 40$<$ to 0.95 (95\% CrI: 0.94, 0.95) in females  80$>$ and coverage probability from 0.92 in females 40$<$ to 0.96 in males 60-69, 70-79: 
\begin{example}
#              Correlation Coverage
# less40F 0.39 (0.37, 0.40)     0.92
# 40-59F  0.78 (0.77, 0.78)     0.95
# 60-69F  0.83 (0.82, 0.83)     0.95
# 70-79F  0.91 (0.90, 0.91)     0.95
# 80plusF 0.95 (0.94, 0.95)     0.95
# less40M 0.51 (0.50, 0.52)     0.93
# 40-59M  0.83 (0.83, 0.84)     0.95
# 60-69M  0.87 (0.86, 0.87)     0.96
# 70-79M  0.92 (0.91, 0.92)     0.96
# 80plusM 0.94 (0.94, 0.94)     0.95
\end{example}

\subsection{Expected number of deaths}

The object \texttt{pois.samples.list} contains 1000 samples from the posterior predictive distribution (\ref{eq:2}), i.e. 1000 samples of the expected number of deaths by age, sex, NUTS3 regions and week, had the pandemic not occurred. We can access the different age-sex groups as follows:
\begin{example}
names(pois.samples.list)
# "F_less40" "F_40_59"  "F_60_69"  "F_70_79"  "F_80plus" 
# "M_less40" "M_40_59"  "M_60_69"  "M_70_79"  "M_80plus"
\end{example}

\noindent where F stands for females and M for males across the different age groups. To get an idea about the structure of the data, we can use the \texttt{head()} function for females 60-69 and check the first 10 samples of excess deaths for the first 6 weeks of 2020 in the 001 region (Torino)
\begin{example}
pois.samples.list$F_60_69 %>% 
select(paste0("V", 1:10), EURO_LABEL, ID_space, year) %>% 
head()
#     V1 V2 V3 V4 V5 V6 V7 V8 V9 V10 EURO_LABEL ID_space year
# 262 22 18 17 15 17 18 13 19 16  17   2020-W01      001 2020
# 263 18 16 21 20 16 23 13 12 23  17   2020-W02      001 2020
# 264 17 23 20 26 12 16 12 19 17  13   2020-W03      001 2020
# 265 13  8 13 30 22 17 16 22 17  19   2020-W04      001 2020
# 266 14 20 15 15 19 18 23 14 12  18   2020-W05      001 2020
# 267 17 14 14  9 14 15 17 19 19  14   2020-W06      001 2020          
\end{example}
	
\noindent  We can also calculate median and 95\% CrI expected number of deaths for this specific age-sex group over the year by NUTS3 region:
\begin{example}
pois.samples.list$F_60_69 %>% 
select(starts_with("V"), "ID_space") %>% 
	group_by(ID_space) %>% 
	summarise_all(sum) %>% 
	rowwise(ID_space) %>% 
	mutate(median = median(c_across(V1:V1000)), 
LL = quantile(c_across(V1:V1000), probs= 0.025), 
UL = quantile(c_across(V1:V1000), probs= 0.975)) %>% 
	select(ID_space, median, LL, UL) %>% 
	head()
		
# A tibble: 107 × 4
# Rowwise:  ID_space
# ID_space median    LL    UL
# <chr>     <dbl> <dbl> <dbl>
# 1 001         814  748   885.
# 2 002          72   55    90 
# 3 003         139  116   167 
# 4 004         212  183.  243.
# 5 005          86   67   107 
# 6 006         178  150   208 
# 7 007          46   33    62 
# 8 008          86   68   107 
# 9 009         113   93   135.
# 10 010         341  301.  380 
# … with 97 more rows
\end{example}
	
\subsection{Excess mortality}
	
The above results can be combined in different ways using the functions \texttt{get2020data()} and \\ \texttt{get2020weeklydata()} to calculate excess mortality. The function \texttt{get2020data()} aggregates over the entire country, NUTS2 regions, sex, age and time resulting in the object \texttt{d}, whereas the function \texttt{get2020weeklydata()} aggregates over the entire country, NUTS2 regions, sex and age but not over time resulting in the object \texttt{d\_week}.
	
\begin{example}
names(d)
# "province" "region"   "country" 
names(d$province)
# "none"   "age"    "sex"    "agesex"
names(d$province$age)
# "40<"   "40-59" "60-69" "70-79" "80+"  
names(d$province$sex)
# "F" "M"
names(d$province$agesex)
# "F40<"   "F40-59" "F60-69" "F70-79" "F80+"   "M40<"   "M40-59" "M60-69" "M70-79" "M80+" 
\end{example}
		
Province stands for the NUTS3 regions (the resolution we used to fit the models), region for the NUTS2 (coarser than NUTS3, appropriate for policy making) and country for the nationwide aggregation. Within these aggregations users can select the option "none" being the total aggregation by age and sex, "age" by sex, "sex" by age  and "agesex" refers to no age and sex aggregation. The objects \texttt{d} and \texttt{d\_week} have similar structure and contain summary statistics for REM and NED and posterior probabilities of a positive REM or NED: 
		
\begin{example}
head(d$province$none)
# Simple feature collection with 6 features and 24 fields
# Geometry type: POLYGON
# Dimension:     XY
# Bounding box:  xmin: 6.626865 ymin: 44.06028 xmax: 9.21355 ymax: 45.95041
# CRS:           +proj=longlat +datum=WGS84
# ID_space SIGLA     DEN_UTS observed population mean.REM median.REM   sd.REM
# 1      001    TO      Torino    32478  2226317.2 21.37414   21.36545 1.116582
# 2      002    VC    Vercelli     3216   168761.7 32.27082   32.15534 3.133142
# 3      003    NO      Novara     5274   364529.4 23.62994   23.71569 2.220801
# 4      004    CN       Cuneo     8716   585479.3 20.16665   20.22069 1.742638
# 5      005    AT        Asti     3745   211437.1 26.37080   26.30691 2.660622
# 6      006    AL Alessandria     7916   416090.6 28.27706   28.21510 2.052881
# LL.REM   UL.REM exceedance.REM median.REM.cat exceedance.REM.cat median.pred
# 1 19.19396 23.50963              1           20%>          (0.95, 1]       26761
# 2 26.51331 38.74179              1           20%>          (0.95, 1]        2434
# 3 19.18442 27.97865              1           20%>          (0.95, 1]        4263
# 4 16.67966 23.54490              1           20%>          (0.95, 1]        7250
# 5 21.19545 31.77340              1           20%>          (0.95, 1]        2965
# 6 24.48449 32.33309              1           20%>          (0.95, 1]        6174
# LL.pred UL.pred mean.NED median.NED    sd.NED   LL.NED   UL.NED exceedance.NED
# 1   26296   27249 5717.153     5717.5 246.42447 5229.975 6182.075              1
# 2    2318    2543  783.265      782.5  57.49419  673.975  898.025              1
# 3    4121    4428 1006.667     1011.0  76.69947  848.925 1153.000              1
# 4    7055    7471 1461.215     1466.0 105.25269 1245.975 1661.075              1
# 5    2842    3092  780.189      780.0  62.31281  654.950  903.000              1
# 6    5982    6360 1743.404     1742.0  98.74248 1556.975 1934.125              1
# median.NED.cat exceedance.NED.cat                       geometry
# 1          1000>          (0.95, 1] POLYGON ((7.859044 45.59758...
# 2    [500, 1000)          (0.95, 1] POLYGON ((8.204465 45.93567...
# 3          1000>          (0.95, 1] POLYGON ((8.496878 45.83934...
# 4          1000>          (0.95, 1] POLYGON ((7.990897 44.82381...
# 5    [500, 1000)          (0.95, 1] POLYGON ((8.046805 45.12815...
# 6          1000>          (0.95, 1] POLYGON ((8.405489 45.20148...
\end{example}

Notice that the object \texttt{d\$province\$none} is a simple feature collection, making mapping it straight-forward.
In Figure~\ref{SpatiotemporalRegions} (plots 1A, 2A and 3A) we show the median posterior of REM for total age and sex at the national, NUTS2 and NUTS3 regional level. For these plots we used the object \texttt{d} with the selection "none" and plot the median REM (\texttt{median.REM}), for example for 3A:
\begin{example}
# prov <- "Foggia"
# d$province$none %>% 
#	filter(DEN_UTS == prov) %>% 
#	select(geometry) %>% 
#	ggplot() +  
#		geom_sf(data = d$province$none, aes(fill = median.REM.cat)) + 
#		geom_sf(fill = NA, col = col.highlight, size = .8) + 
#		scale_fill_manual(values=colors, name = "", drop=FALSE) + 
#       theme_light() + ggtitle(paste0("3A. NUTS3 regions: ", prov)) 
\end{example}
			
\noindent Overall, the REM in Italy during 2020 was between 15-20\%, Figure~\ref{SpatiotemporalRegions}, panel 1A. When the higher geographical resolution is assessed, it is revealed that north and in particular Lombardia is the region hit the worst, with the REM exceeding $20\%$, Figure~\ref{SpatiotemporalRegions}, panels 1B and 1C. Figure~\ref{PosteriorProb} shows a measure of uncertainty of the REM, now for the different age groups and both sexes (selection "age"). The probability of a positive excess (\texttt{exceedance.REM}) in older people is larger than 0.95 almost everywhere, Figure~\ref{PosteriorProb}.
		
Panels 1B and 1C of Figure~\ref{SpatiotemporalRegions} show the median temporal nationwide excess together with 95\% CrI by sex after aggregating the different age groups (using \texttt{d\_week} and the "sex" selection). We observe a clear first pandemic wave during March and May and a second one during mid October and December in 2020. During the first pandemic wave, there were weeks when the median REM reached almost 100\% in males, Figure~\ref{SpatiotemporalRegions}.
		
Panels 2B, 2C, 3B and 3C of Figure~\ref{SpatiotemporalRegions} show the median spatio-temporal excess together with 95\% CrI by sex after aggregating over the different age groups. Panels 2B and 2C highlight the region of Puglia, where during the first wave of the pandemic experienced a positive, similar with the nationwide excess. When we increase the spatial resolution in panels 3B and 3C, we highlight the province of Foggia, where there was insufficient evidence of a positive excess during the first wave, but strong during the second. 

\subsection{Shiny Web-Application}
To be able to effectively examine and communicate the different aggregation levels of the output of our modelling framework, we have also developed a Shiny Web-Application (WebApp), Figure~\ref{shinyapp}. The WebApp provides spatial, temporal and spatio-temporal analysis tabs, and within each tab there are plots and summary statistics for the level of aggregation selected from the drop-down menu. Users can select across different variables (REM or NED), statistics (median or posterior probability), sex (males, females or both), age group ($40<$, $40-59$, $60-69$, $70-79$, $80>$ and all) and different geographical level (national, NUTS2 or NUTS3 regions). Summary statistics for each area are available and they are displayed in a pop-up window, which is activated by clicking on the area of interest. In addition, graphical pop-ups are provided to show the weekly estimates for each area with the \texttt{leafpop} \texttt{R}-package \citep{leafpop}, in the spatio-temporal analysis tab. \\
The WebApp that we have developed is hosted at \url{http://atlasmortalidad.uclm.es/italyexcess/}.  

\begin{widefigure}[H]
	\centering
	\includegraphics{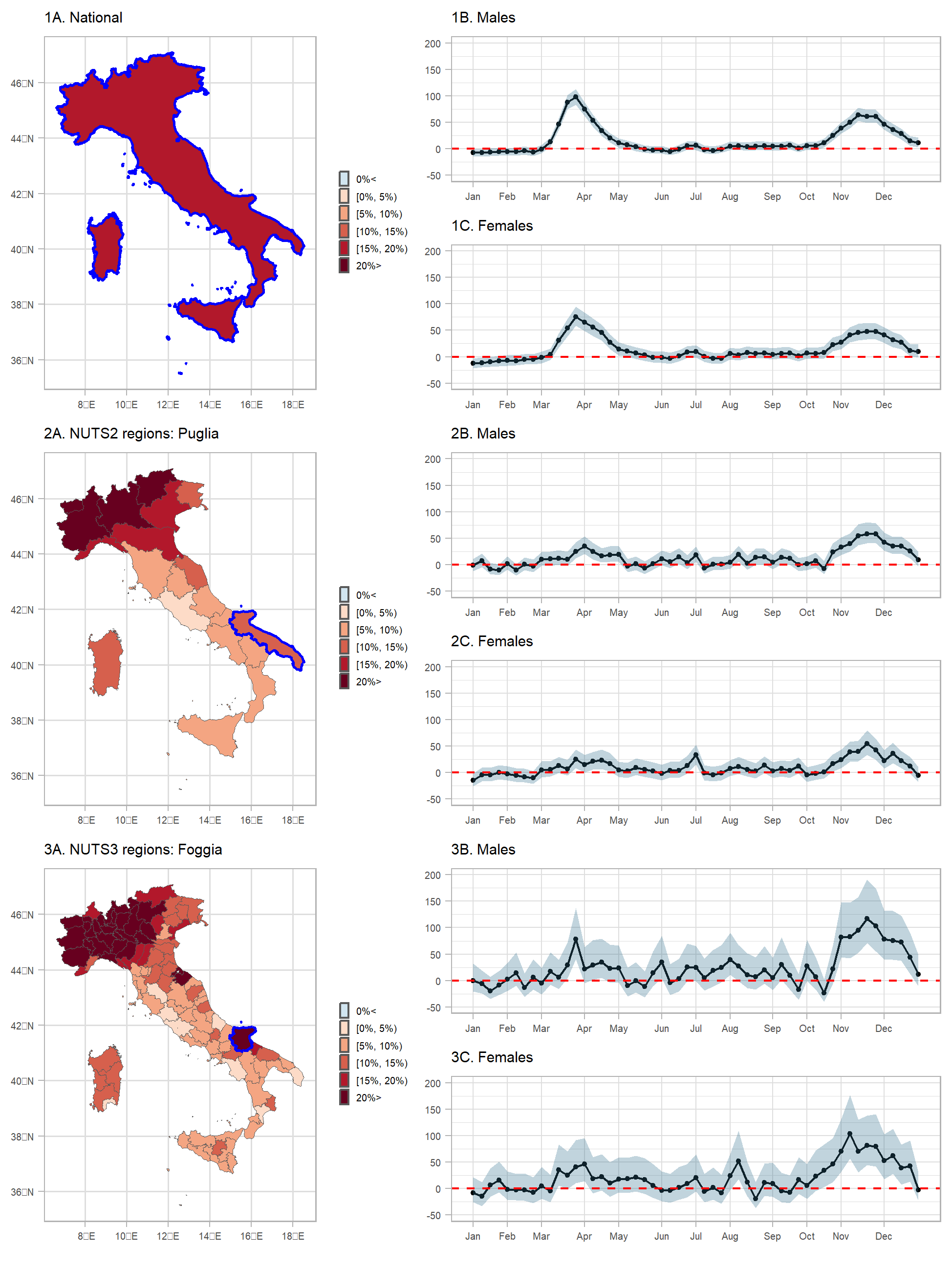}
	\caption{Median relative excess mortality by spatial region during 2020: 1A. nationwide, 2A. NUTS2 and 3A. NUTS3 level, and weekly median relative excess mortality and 95\% Credible Intervals (95\% probability that the true value lies within this interval) by sex during 2020 in: 1B. males nationwide, 1C. females nationwide, 2B. males in Puglia, 2C. females in Puglia, 3B. males in Foggia and 3C. females in Foggia.}
	\label{SpatiotemporalRegions}
\end{widefigure}	
		
\begin{widefigure}[H]
	\begin{flushright}
		\includegraphics{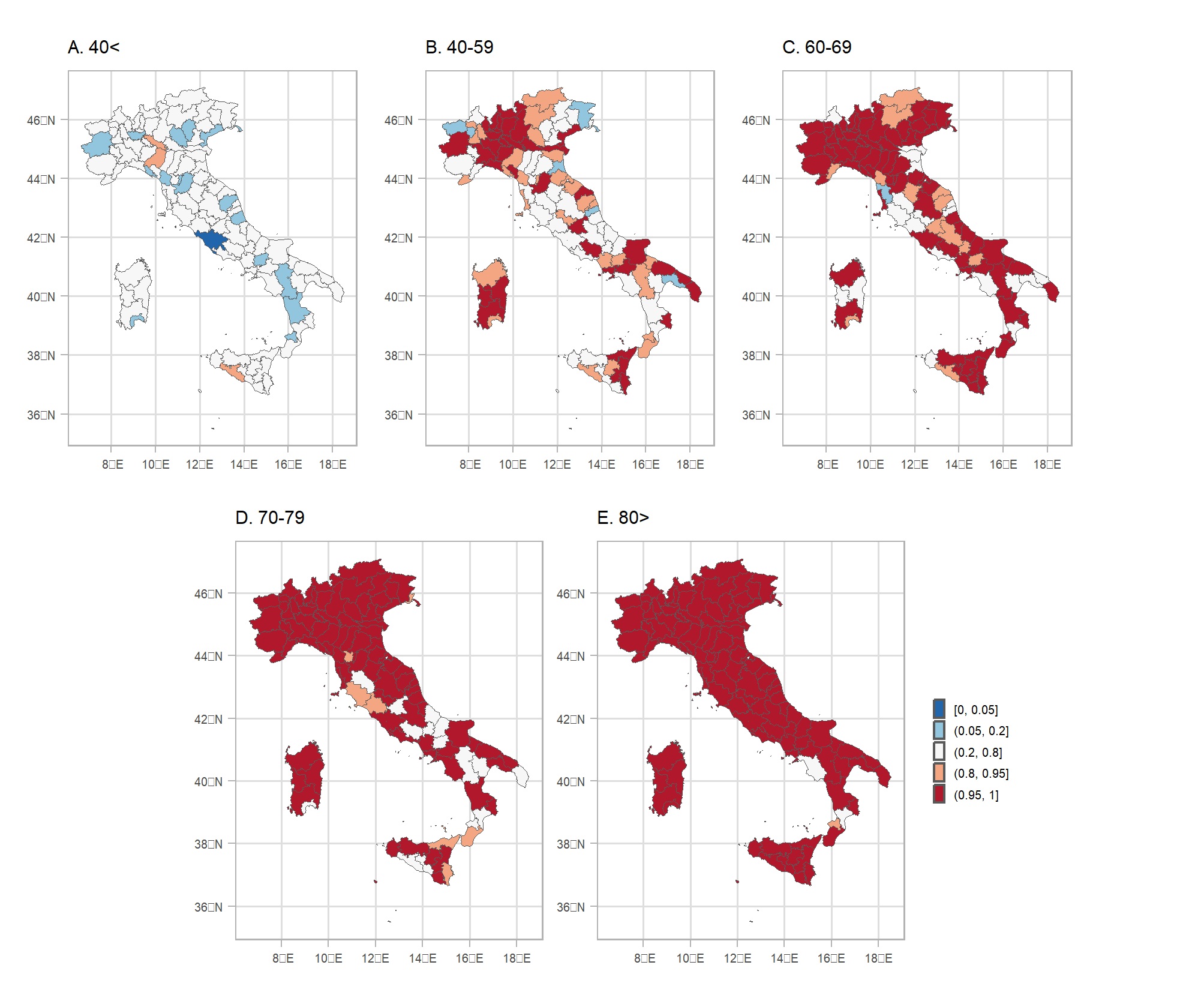}
	\end{flushright}
	\caption{Posterior probability that the relative excess mortality is positive for both sexes during 2020 by age group and NUTS3 region. }
	\label{PosteriorProb}
\end{widefigure}
		
\begin{widefigure}[!t]
	\begin{center}
		\includegraphics[width=0.9\textwidth]{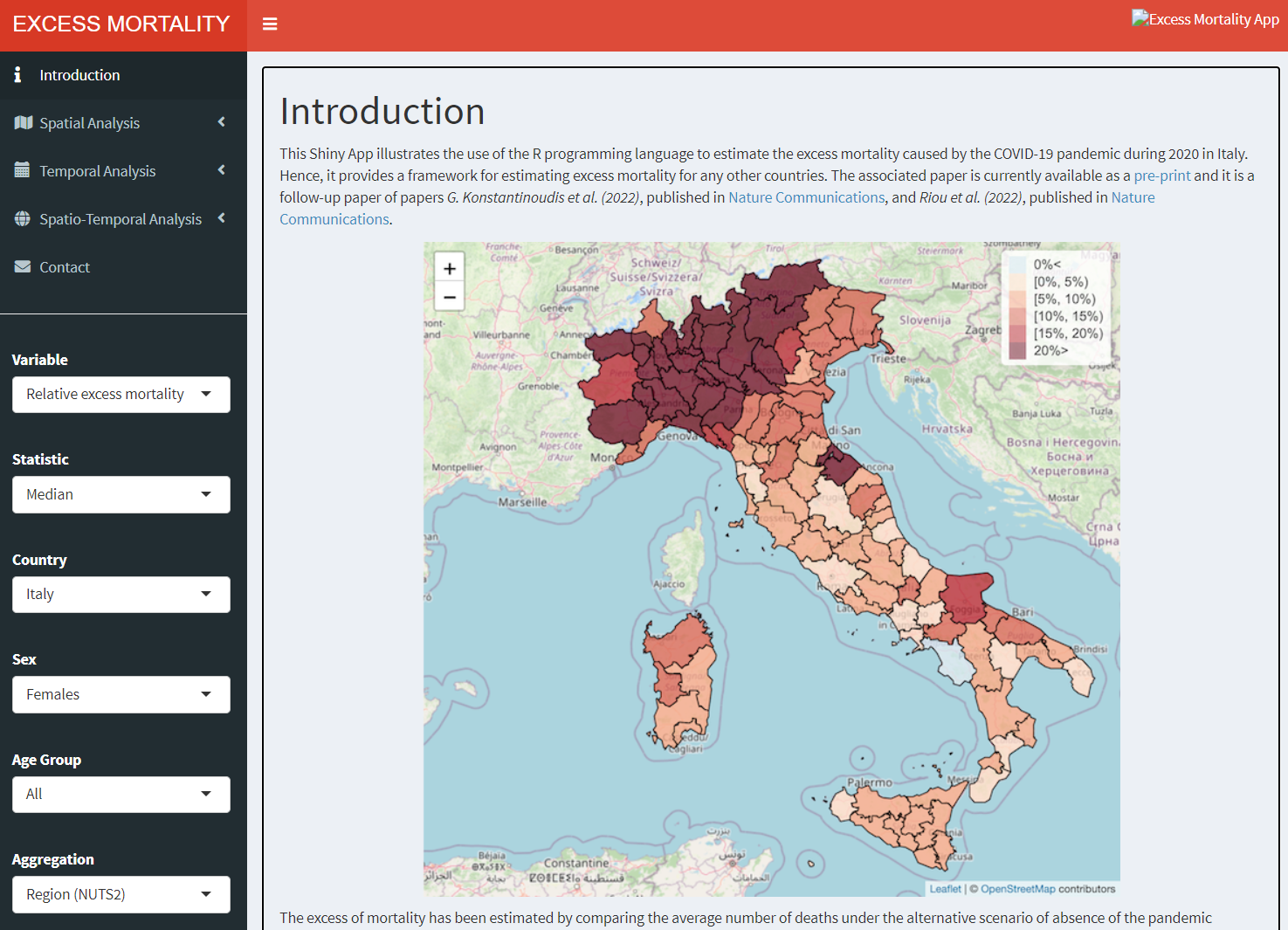}
	\end{center}
	\caption{Shiny App developed to navigate through the excess mortality estimates during 2020 in Italy across the different aggregations.}
	\label{shinyapp}
\end{widefigure}
		
\section{Summary}
This tutorial provides a detailed implementation of the framework followed previously to calculate excess mortality during the COVID-19 pandemic in 5 European regions \citep{konstantinoudis2021regional}. The main model used here is slightly modified based on updated results \citep{riou2023direct}. We have proposed a Bayesian framework for estimating excess mortality and shown how to use \texttt{R} and \texttt{INLA} to retrieve fast and accurate estimates of the excess mortality. The proposed framework also allows combining different models and presenting the results in any age, sex, spatial and temporal aggregation desired. We have given a practical example of how to use the proposed framework modelling the excess mortality during the 2020 COVID-19 pandemic in Italy at small area level. We also developed a Shiny App to effectively communicate the results. This framework can be extended to monitor mortality from other extreme events for instance natural hazards such as hurricanes \citep{acosta2020flexible}. Other extensions include different ways of modelling the younger age groups to increase the predictive ability, for instance using a zero-inflated Poisson. All the above make the proposed framework particularly flexible, powerful, generalisable and appealing for online monitoring of the pandemic burden and timely policy making. 
			
\section*{Acknowledgements}
		
All authors acknowledge infrastructure support for the Department of Epidemiology and Biostatistics provided by the NIHR Imperial Biomedical Research Centre (BRC). The authors also acknowledge infrastructure support for the domain of the shiny app from the University of Castilla-La Mancha.
		
G.K. is supported by an MRC Skills Development Fellowship [MR/T025352/1]. M.B. is supported by a National Institutes of Health, grant number [R01HD092580-01A1]. Infrastructure support for this research was provided by the National Institute for Health Research Imperial Biomedical Research Centre (BRC). The work was partly supported by the MRC Centre for Environment and Health, which is funded by the Medical Research Council (MR/S019669/1, 2019-2024). V.G.R. is supported by grant SBPLY/17/180501/000491 and SBPLY/21/180501/000241, funded by Consejer\'ia de Educaci\'on, Cultura y Deportes (JCCM, Spain) and FEDER, and grant PID2019-106341GB-I00, funded by Ministerio de Ciencia e Innovaci\'on (Spain). We thank Univesidad de Castilla-La Mancha for hosting the server on which the Shiny App is running.

\section*{Author contributions}
		
V.G.R. conceived the study. M.B. supervised the study. G.K. developed the initial study protocol and discussed it with M.B., M.C., M.P. and G.B.. G.K. developed the statistical model, prepared the population and covariate data and led the acquisition of mortality data. M.C. and V.G.R. validated and modified accordingly the code. G.K. ran the analysis. G.K., V.G.R., M.B., M.C. and M.P wrote the initial draft and all the authors contributed in modifying the paper and critically interpreting the results. V.G.R. developed the Shiny app. All authors read and approved the final version for publication.
		
\section*{Competing interests}
The authors declare no competing interests.
		
\section*{Ethics}
The study is about secondary, aggregate anonymised data so no additional ethical permission is required.
		
\bibliographystyle{unsrt}
\bibliography{RJreferences}

\address{Garyfallos Konstantinoudis\\
	MRC Centre for Environment and Health, \\ Imperial College London, \\ St Mary's Campus, Praed St, \\ W2 1NY, London\\
	United Kingdom \\}
\email{g.konstantinoudis@imperial.ac.uk}
		
\address{Virgilio G\'{o}mez-Rubio\\
	Departamento de Matem\'aticas, Escuela T\'ecnica Superior de Ingenieros Industriales-Albacete, \\ Universidad de Castilla-La Mancha, \\ Albacete, Spain \\}
\email{virgilio.gomez@uclm.es}
		
\address{Michela Cameletti\\
	Department of Economics, University of Bergamo, \\ Bergamo, Italy\\
}
\email{michela.cameletti@unibg.it}
		
\address{Monica Pirani\\
	MRC Centre for Environment and Health, \\ Imperial College London, \\ St Mary's Campus, Praed St, \\ W2 1NY, London\\
	United Kingdom \\}
\email{monica.pirani@imperial.ac.uk}
		
\address{Gianluca Baio\\
	Department of Statistical Sciences, \\ University College London, \\ United Kingdom\\ }
\email{g.baio@ucl.ac.uk}
		
\address{Marta Blangiardo\\
	MRC Centre for Environment and Health, \\ Imperial College London, \\ St Mary's Campus, Praed St, \\ W2 1NY, London\\
	United Kingdom \\}
\email{m.blangiardo@imperial.ac.uk}